\documentclass[aps,prb,epsf,eqsecnum,twocolumn,showpacs]{revtex4}
\usepackage{graphicx}
\usepackage{amsmath}
\begin{document}
\title{Upper and lower bounds for the large polaron dispersion in $D=1,2,3$ dimensions}
\author{Bernd Gerlach}
\affiliation{Institut f\"ur Physik, Universit\"at Dortmund, 44221 Dortmund, Germany}
\author{Mikhail A. Smondyrev}
\affiliation{N. N. Bogoliubov Laboratory of Theoretical Physics, Joint Institute for Nuclear Research, 141980 Dubna, Russia} %++
\date{\today}

\begin{abstract}
Numerical results for the polaron dispersion
are presented for an arbitrary number of space dimensions. Upper
and lower bounds are calculated for the dispersion curves. They
are rather close to each other in the cases of small
electron-phonon couplings usual for real polar materials. To describe the dispersion
in other materials, we suggest a simple fitting formula
which can be applied at intermediate values of the Fr\"ohlich electron-phonon
coupling constant. Its validity is
approved by the comparison with direct
calculations and previously obtained results. This makes our
results not only reliable and highly accurate but also easy
reproducible.
\end{abstract}

\maketitle

%\narrowtext

\section{Introduction}

The electron-phonon interaction influences the properties of charge carriers
(electrons) in polar semiconductors or ionic crystals. An electron polarizes a medium and,
being surrounded by a cloud of virtual phonons, is captured by a self-induced potential
which can move in a material. Such a quasiparticle is called a {\it polaron}.
The larger the value of the Fr\"ohlich electron-phonon coupling constant $\alpha$,
the more pronounced are polaron effects. In particular, the electron-phonon interaction
results in an electron binding energy, in a renormalization of its mass and in a nonparabolic
energy-momentum dependence.

In the present paper, we study the large polaron dispersion law
that is the dependence of the polaron ground-state energy
$E(\alpha, Q)$ on the total polaron momentum $Q$.
The technological progress in
man-made structures has caused a rapidly increasing literature on systems of reduced
dimensionality, in particular, on the effects of electron-phonon
interaction in quantum wells, wires and dots. Thus,  the
dimensionality $D$ of space in our study may be different: $D=1,2,3$.

Here and in what follows all quantities are dimensionless, the energy, mass and
length units being $\hbar\omega$, $m$ and $\sqrt{\hbar/2m\omega}$, respectively,
where $\omega$ is the longitudal optical (LO) phonon frequency
and $m$ is the free electron band mass. For instance, the polaron kinetic energy $P^2/2m_{pol}$
at small momentum $P$ is written down as $Q^2/m(\alpha)$ in our dimensionless units
where $m_{pol}$ is the polaron effective mass and $m(\alpha)=m_{pol}/m$.

The literature on polaron is enormous but it concerns mostly the
energy $E(\alpha, 0)$ of the bulk ($D=3$) polaron at rest.
Nevertheless, the first results were obtained on a bulk polaron dispersion law
even in a very early paper by Fr\"ohlich, Pelzer and Zienau \cite{zienau} where the authors used
the first order of the Brillouin-Wigner perturbation theory. Later Whitfield and Puff
\cite{whit} suggested an improved version of the polaron energy-momentum relation in the weak-coupling regime.
For these and other early papers see also the review article by Appel\cite{appel} where qualitative
considerations on the behavior of the dispersion curve were given.
The results obtained demonstrate that the bulk polaron energy-momentum relation is quadratic
for small $Q$ but then bends over and becomes horizontal
when the energy approaches the continuum edge $E_c$ which is reached at some finite value $Q_c$
of the polaron momentum. At this momentum the moving polaron energy $E_c=E(\alpha,Q_c)$ exceeds the ground-state
energy $E(\alpha,0)$ of the polaron at rest exactly by the energy of a free phonon $\hbar
\omega$ (which is just unity in our notation).

Below the continuum edge $E_c=E(\alpha, 0)+1$ (that is at $Q < Q_c$)
the ground-state energy $E(\alpha, Q)$ is an isolated and well defined eigenvalue.
There are some important rigorous results concerning the
properties of the dispersion $E(\alpha, Q)$ (see
Ref.~\cite{lowen,spohn})

\begin{itemize}
\item $E(\alpha, Q)$ is a real analytic function of $\alpha$ and
$Q$ for $0 \leq \alpha < \infty, Q^2 \leq 1$. The former
constraint on $Q$ can be removed totally for $D=1, 2$. For
$D=3$ the domain of $Q$ can be extended up to a finite value
$Q_c$, where the energy reaches the continuum edge.
\item $E(\alpha, Q)$ decreases with $\alpha$ and increases with
$Q$ below the continuum edge.
\item The inequality $E(\alpha, 0) <  E(\alpha, Q)$ holds for $Q \neq 0$ and $0 \leq \alpha < \infty$.
\item The upper bound is given by the inequality \begin{equation}\label{rigorous}
E(\alpha, Q) \leq \min [ E(\alpha, 0) + Q^2, E(\alpha, 0)+1].
\end{equation}
\end{itemize}

The simplest and seemingly the most natural way to describe the
dispersion is the first order of the Raleigh-Schr\"odinger
perturbation theory (RSPT). For $D=1,2,3$ it results in the
well-known formulae:
\begin{subequations} \label{RSPT}
\begin{eqnarray}
E(\alpha, Q) &=& Q^2 - \alpha {\pi \over 2\sqrt{1-Q^2}}\ {\rm
for}\ D=1, \label{RSPT-1}  \\
E(\alpha, Q) &=& Q^2 - \alpha K(Q^2) \ {\rm for}\ D=2 , \label{RSPT-2} \\
E(\alpha, Q) &=& Q^2 - \alpha {\arcsin Q \over Q} \ {\rm
for}\ D=3 . \label{RSPT-3}
\end{eqnarray}
\end{subequations}
Here $K(m)$ is the complete elliptic integral of the first kind.

The results of the RSPT contradict the rigorous properties of the
dispersion. For instance, all three functions of Eq.~(\ref{RSPT})
have maxima at some momenta $Q_m < 1$ so they are decreasing functions
in the region $Q_m < Q < 1$. Moreover, the expressions for $D=1, 2$
diverge at $Q=1$, so the RSPT  evidently fails to
work near this value. At small $Q \ll 1$ RSPT leads however to
correct (to the first order in $\alpha$) parabolic functions of
the effective mass approximation.

After the above cited paper\cite{zienau} different variational approaches
were developed to calculate the polaron dispersion. The paper by Lee, Low and Pines \cite{lee}
should be mentioned
among earlier articles on polarons. For sufficiently large $Q$
their variational result is weaker than the nonanalytical upper bound
(\ref{rigorous}), the dispersion curve intersects the continuum
edge and the deviation from the correct result is of the order
$\alpha^0$, which does not provide the necessary accuracy of
calculations.

More advanced variational calculations were performed by
Larsen\cite{larsen}, and by Warmenbol, Peeters and Devreese
\cite{warm2,warmenbol}. These and some other papers will be
discussed in more detail later.

Naturally, only upper bounds could be obtained with the variational methods of these papers.
It is much more difficult to derive a lower bound, and it was found
by Lieb and Yamazaki \cite{lieb} but only for $Q=0$. The
general procedure to obtain the $Q-$dependent lower bound was developed by
Gerlach and Kalina \cite{gerl3}.

In the present paper, we combine the variational upper bound
obtained with the expansion of the trial wave function in numbers
of virtual phonons with the lower bound of Ref.~\cite{gerl3}. In
principle, our variational upper bound could lead to exact
solutions but in practice we have to cut the expansion and work
with an approximation obtained this way. For small
values of the electron-phonon coupling constant $\alpha$ which
are common for most of the polar materials the corridor between
these two estimates is very narrow, so that we can pretend to
finding numerically exact solutions.

The lower bound gives too poor results for intermediate values $\alpha \sim 1$.
Besides the huge numerical job does not allow us to
reach the necessary accuracy with the upper bound. To overcome
these difficulties, we suggest simple fitting formulas
 to calculate the polaron dispersions in different dimensions
for intermediate values of the coupling constant. This makes our results
easy reproducible and reliable, which is demonstrated while comparing them with
these by other authors.

\section{Basic Equations}

The starting point is a Hamiltonian of Fr\"ohlich type in the form
which makes use of translation invariance to perform a projection
onto a subspace of fixed total polaron momentum:

\begin{eqnarray} \label{H}
H({\mathbf{Q}}):&=& \left({\mathbf{Q}}-\sum_{\mathbf{k}}{\mathbf
k}a^{\dag}_{\mathbf{k}}a_{\mathbf{k}}\right)^{2} +
\sum_{\mathbf{k}}a^{\dag}_{\mathbf{k}}a_{\mathbf{k}} \nonumber \\
&&
+\sum_{\mathbf{k}}(g_{\mathbf{k}}a_{\mathbf{k}}+g^{*}_{\mathbf{k}}a^{\dag}_{\mathbf{k}})
\ \ .
\end{eqnarray}

\noindent The electron coordinates were eliminated with the
well-known Lee-Low-Pines canonical transformation \cite{lee}.
This reflects the conservation of the total polaron momentum
${\mathbf{Q}}$ which is a c-number in Eq.~(\ref{H}). Here
${\mathbf{k}}, g_{\mathbf{k}}, a_{\mathbf{k}}$, and
$a^{\dag}_{\mathbf{k}}$ are the wave vector, coupling function,
and the annihilation- and creation operator of the phonon under
consideration. The coupling function is defined as

\begin{equation}  \label{g}
g_{\mathbf{k}}=\sqrt{\frac{\pi\alpha{\xi}_D}{Vk^{D-1}}} \ \ ,
\end{equation}
where $V$ is the quantization volume and ${\xi}_{D}$ is a number. For $D=3$ and
${\xi}_{3}=4$ or $D=2$ and ${\xi}_{2}=2$ one recovers well-known
models (see, e.g. Ref. \cite{peet1}). At first glance, the case
$D=1$ has to be excepted as, according to Ref. \cite{peet1},
${\xi}_1$ diverges. Nevertheless, the coupling $g_{\mathbf{k}}$ is
physically interesting for $D=1$ and finite ${\xi}_1$ - either in
the sense of a regularized version of a polaron model, as
discussed in Ref. \cite{peet2}, or in the sense of an effective
model within the theory of the bulk ($D=3$) free polaron
model \cite{gerl2}. We choose ${\xi}_1=1$ without loss of
generality.

Summations in final formulae are replaced by integrations
following the conventional rule:
\begin{equation}
\sum_{\mathbf{k}}\mid g_{\mathbf{k}}\mid^2 F({\mathbf{k}}) = {\pi
\alpha \xi_D \over (2\pi)^D}\int {d {\mathbf{k}}\over
k^{D-1}}F({\mathbf{k}}).
\end{equation}

The scheme to calculate the upper bound for $E(\alpha,{\bf{Q}})$
was presented in Ref. \cite{gerl3}. The variational principle of
Ritz was used. Choosing an adjustable, normalized wave function
$\mid \Phi
>$ and calculating the minimum of $<\Phi \mid H({\mathbf{Q}})
\mid \Phi >$, one gets an upper bound $z\equiv z(\alpha, Q)$ to
$E(\alpha,Q)$. Our trial function is of the type
\begin{eqnarray}  \label{ans1}
&&\mid \Phi >: = C\left[ \mid
0>-\sum_{\mathbf{k}}g_{\mathbf{k}}B_{\mathbf{k}}a^{\dag}_{\mathbf{k}}\mid
0> \right. \\
&+&\left.
\sum_{\mathbf{k},\mathbf{k'}}g_{\mathbf{k}}g_{\mathbf{k'}}B_{\mathbf{k},\mathbf{k'}}a^{\dag}_{\mathbf{k}}a^{\dag}_{\mathbf{k'}}\mid
     0> - \right. \nonumber \\
     &-&\left. \sum_{\mathbf{k},\mathbf{k'},\mathbf{k''}}g_{\mathbf{k}}g_{\mathbf{k'}}g_{\mathbf{k''}}B_{\mathbf{k},\mathbf{k'},\mathbf{k''}}a^{\dag}_{\mathbf{k}}a^{\dag}_{\mathbf{k'}}
     a^{\dag}_{\mathbf{k''}}\mid 0> + \ldots \right] \nonumber ,
\end{eqnarray}
the two- and three-phonon amplitudes being totally symmetrical
functions. The mean value $<\Phi \mid H({\mathbf{Q}}) \mid \Phi>$
can be readily calculated from which we obtain the minimizing
equations for the amplitudes. The more phonon amplitudes are
included, the better is the upper bound. In principle, one can
include an arbitrary number of phonons arriving at a subsequently
large number of equations for the corresponding amplitudes.

In practice, one has to cut expansion (\ref{ans1}). For example,
restricting ourselves to the one-phonon amplitude we arrive at
the Brillouin-Wigner perturbation result in the first order in
$\alpha$:
\begin{equation}
z= Q^2 - \sum_{\mathbf k} {|g_{\mathbf k}|^2 \over (\mathbf{
Q-k})^2 +1 -z}. \label{1phonon}\end{equation}

If we take into account the two-phonon contribution, the system
of subsequent equations takes the form:
\begin{equation}
z=Q^2-\sum_{\mathbf{k}}|g_{\mathbf{k}}|^2 B_{\mathbf{k}}
\label{2phonon-1}
\end{equation}
and
\begin{equation}  \label{B1-a}
\left[ (\mathbf{Q-k})^{2}+1-z \right] B_{\mathbf{k}} =
1+2\sum_{\mathbf{k'}}|g_{\mathbf{k'}}|^2 B_{{\mathbf{k}},{\mathbf{k'}}},  %++
\end{equation}

\begin{equation}  \label{B2-a}
2\left[ (\mathbf{ Q -k - k'})^{2}+2-z
\right]B_{\mathbf{k},\mathbf{k'}} =
B_{\mathbf{k}}+B_{\mathbf{k'}}. \end{equation}

The quantity $B_{\mathbf{ k, k'}}$ can be readily found from
Eq.~(\ref{B2-a}). Inserting it into Eq.~(\ref{B1-a}) and defining
the function
\begin{eqnarray}
d({\mathbf x}; z) = \sum_{\mathbf{k'}}{|g_{\mathbf{k'}}|^2 \over
(\mathbf{ x -k'})^2+2-z} \label{D-a}\end{eqnarray} we arrive at
the equation for the quantity $B_{\mathbf k}$:
\begin{eqnarray}
&& \left[ (\mathbf{ Q- k})^{2}+1-z -d(\mathbf{ Q-k}; z)\right]
B_{\mathbf{k}}  =  \nonumber \\
&& 1+\sum_{\mathbf{k'}}{|g_{\mathbf{k'}}|^2 B_{\mathbf k'}\over (\mathbf{ Q-k-k'})^2+2-z }.  %++
\label{B1-b}\end{eqnarray}

To keep the necessary accuracy of the solution, we iterate it
twice and truncate the series:
\begin{eqnarray}
B_{\mathbf k} &=& {1\over N(\mathbf{ Q-k}; z) }  \\ && +
\sum_{\mathbf{k'}}{|g_{\mathbf{k'}}|^2 \over N(\mathbf{ Q-k};
z)[(\mathbf{ Q-k-k'})^2+2-z] }.\nonumber
\label{B1-c}\end{eqnarray} where
\begin{eqnarray}
N({\mathbf x}; z)= x^2+1-z-d({\mathbf x}; z).
\label{propagator}\end{eqnarray} Inserting Eq.~(\ref{B1-c}) into
Eq.~(\ref{2phonon-1}) we obtain the upper bound:
\begin{eqnarray}
&& z = Q^2 - \sum_{\mathbf{k}}{|g_{\mathbf{k}}|^2 \over N(\mathbf{
Q-k};z)} \\
&& - \sum_{\mathbf{k,k'}}{ |g_{\mathbf{k}}|^2 |g_{\mathbf{k'}}|^2
\over N(\mathbf{Q-k};z)[(\mathbf{ Q-k-k'})^2+2-z]N(\mathbf{
Q-k'};z)}.\nonumber \label{2phonons}\end{eqnarray}

Recall that the dispersion curve starts at the minimal value
$E(\alpha, 0)$ and reaches its maximal value at the continuum edge.
The exact expression for the latter is as follows:
\begin{equation}
E_c=E(\alpha,0)+1 .
\end{equation}
In our variational estimates this formula is modified:
\begin{equation}\label{eqn}
E_{c, n\, {\rm ph}}=E_{(n-1)\, {\rm ph}}(\alpha,0)+1 ,
\end{equation}
where the lower indices show the number of phonon amplitudes
taken into account. Here $ n \geq 1$ and $E_{0\, {\rm
ph}}(\alpha,0)=0$ (no electron-phonon interaction). The
interpretation of Eq.~(\ref{eqn}) is clear: as we work with
a fixed number of phonon amplitudes and one phonon becomes free
at the continuum edge, the number of phonons which contribute to
the polaron state decreases by one.

Therefore, in the one-phonon approximation we obtain for the
continuum edge a rather trivial and poor estimate $E_{c, 1\, {\rm
ph}}=1$. If a three-phonon amplitude is included, we describe
$z(\alpha, Q)$ with high accuracy (e.g., correct up to terms of
order ${\alpha}^2$ near the continuum edge and up to terms of order ${\alpha}^3$
for the small-$Q$ behaviour). This would be enough for reasonable estimates in the
weak coupling regime. The only problem is that the numerical job becomes time
consuming and could be done only for the case $D=1$ (see Ref.
\cite{kalina}).

To conclude this section we mention a qualitatively different behavior
of the dispersion curves for $D=1,2$ and $D=3$. The continuum edge is
an asymptote for $D=1,2$, and is
approximated from below as $Q\to \infty$, whereas for $D=3$ the
dispersion does meet the edge at a finite value of $Q=Q_c$, which
depends on $\alpha$ (see Refs. \cite{whit,spo2,gerl3}). This
distinction between the low-dimensional and bulk polaron dispersions
is explained by properties of the interaction potential and
may be readily understood.
The total polaron energy $z$ is presented in the r.h.s. of Eq. (\ref{1phonon}) as a sum of a positive
free polaron kinetic energy $Q^2$ and a negative interaction energy proportional to $\alpha$.
At $D=1, 2$ the latter becomes infinitely large when the total energy $z$ approaches the continuum edge
(unity in this approximation) because the integral in Eq. (\ref{1phonon}) diverges.
To compensate this and to keep the total energy finite,
the kinetic energy (and the polaron momentum) should tend to
infinity as well. This is not the case in 3D where the interaction energy stays finite at the continuum edge and so does
the limiting value $Q_c$ of the total polaron momentum.

\section{The 1D Case}

With only one-phonon exchange taken into account formulae
(\ref{g}) and (\ref{1phonon}) result in the following equation for
the upper-bound $z(\alpha,{\mathbf Q})$ at $D=1$:
\begin{equation}
z=Q^2 -\alpha {\pi\over 2\sqrt{1-z}}\ .
\label{1d-1phonon}\end{equation} If we put $z=Q^2$ in the r.h.s.
of Eq.~(\ref{1d-1phonon}), we immediately arrive at the expression
for the first order of RSPT given by Eq.~(\ref{RSPT-1}).

An upper bound obtained with the
inclusion of the two-phonon amplitude can also be presented in a
closed analytical form:
\begin{eqnarray}
z & = & Q^2 -{\alpha \pi\over 2}{1\over  \sqrt{1-z-d(z)}}- \nonumber \\
&& - \left( {\alpha \pi\over 2}\right)^2\left[ 1+2 \sqrt{1-z-d(z)
\over 2-z}\right] \cdot \nonumber \\
&&\cdot {1\over Q^2 +[\sqrt{2-z}+2\sqrt{1-z-d(z)}]^2}\ ,
\label{2ph-final}\end{eqnarray} where
\begin{equation}
d(z)= \alpha {\pi \over 2 \sqrt{2-z}}\ .
\end{equation}

\begin{figure}[h]
%   \hspace*{-1.0cm}
%   \epsfysize=2.7in
   \includegraphics[width=3.7in]{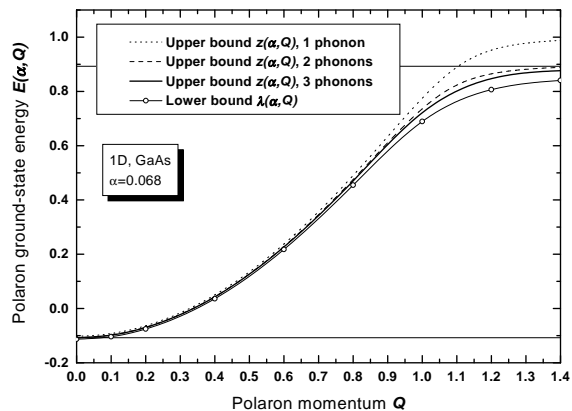}
%   \vspace*{-2.5cm}
\caption{Dispersion of 1D polaron in GaAs}
\label{fig1}
\end{figure}

General formulae and some numerical results for a three-phonon
exchange as well as a lower bound can be found in Ref.
\cite{kalina}. Here they are presented in Fig.~\ref{fig1} for a
very small value of the coupling constant $\alpha=0.068$
corresponding to a real material (GaAs).

\section{Fitting formula}

To give an idea of typical values
of the Fr\"ohlich coupling constant for other materials, we present here some experimental data from Ref. \cite{karth}:
$\alpha = 0.022$ for InSb, $\alpha= 0.123$ for AlAs, $\alpha = 0.65$ for ZnS,
$\alpha = 1.84$ for AgCl. It is well known that intermediate
values $\alpha \sim 1$ and even larger can be tackled within conventional Raleigh-Schr\"odinger perturbations.
This becomes possible because the coefficients $e_n$ in the expansion $E(\alpha,0) = \sum_n e_n \alpha^n$
of the polaron ground-state energy decrease
very fast with $n$ so that the expansion is performed, roughly speaking, in powers of $\alpha/10$ or so.
One can notice this tendency (although it is not proved rigorously) looking at Eq. (\ref{eq-perturb}).
The famous Feynman approximation \cite{feynman} clearly
demonstrates this property as it follows from our calculations \cite{smond2} of first twelve (sic!)
coefficients of the weak-coupling expansion for the 3D-polaron. Besides, the radius of convergence $R_\alpha$
of the RSPT for the bulk polaron ground-state energy was estimated in a number of papers:
Larsen \cite{larsen2} found $R_\alpha \sim 6.2 \div 6.5$,
Klochikhin \cite{klochikhin} argued that $R_\alpha \sim 3.4$, our estimate \cite{smond2} gave $R_\alpha \sim 6.9$.

Anyway, this explains why the weak coupling expansion can be applied
for intermediate values $\alpha  \sim 1$ or even larger.
On the other hand, our lower bound gives very poor results for the intermediate coupling.
In addition, the two-phonon approximation works not so
well at these values of the coupling constant $\alpha$. The reason is that
the edge point $E(\alpha,0)+1$ is then calculated within the
one-phonon approximation, which is certainly not enough to reach appropriate accuracy.

Taking as an example 1D-polaron, $E_{1 ph}(\alpha,0)$ is defined
as $z$ of Eq.~(\ref{1d-1phonon}) at $Q=0$:
\begin{equation}\label{1phon}
  E_{1 ph}(\alpha,0)\sqrt{1-E_{1 ph}(\alpha,0)}=-\alpha{\pi\over 2}.
\end{equation}
Expanding $E_{1 ph}(\alpha,0)$ in powers of $\pi\alpha /2$ we obtain
$E_{1 ph}(\alpha,0) = - (\alpha \pi/2)+ (1/2)(\alpha \pi/2)^2 +
{\cal{O}}(\alpha)^3$. Here the second order coefficient is
positive and equals one-half, while in the correct result \cite{peet2}
$$E_{1 ph}(\alpha,0) = - \left( {\alpha \pi \over 2} \right)- \left({3\sqrt{2}\over 4}-1\right)\left({\alpha
\pi \over 2}\right)^2 + {\cal{O}}(\alpha)^3$$
 it is small (-0.06) and negative. From the point of view of
diagrammatic technique for polarons\cite{smond} it means that
only the disconnected Feynman diagram is taken into account in the one-phonon approximation.
In this approximation the
contribution of two connected diagrams is skipped which otherwise would compensate rather
a large positive coefficient to give a small negative
residue -0.06. An analogous fine tuning happens for flat ($D=2$) and
bulk ($D=3$) polarons as well.

This explains why one needs to
include three-phonon amplitudes to describe not only weak
couplings but also intermediate values $\alpha \sim 1$. However, as was mentioned earlier,
three-phonon calculations can be performed only in $D=1$,
and the numerical job becomes enormous in other spatial dimensions.
So we propose now a simple formula to fit the exact
dispersion curves. We demonstrate its validity for $D=1$ by the
comparison with the three-phonon approximation. In other sections
of the current paper the fitting formula will be used instead of the absent
numerical three-phonon calculations.

Let us stress that we are not going to proceed to the strong-coupling limit. We are still
dealing with the weak couplings and our goal is to restore what is
lost in approximations with partial summation of the perturbation series, that is to regain
a possibility to use the weak-coupling results for the intermediate values of $\alpha$.

We start with the first term of the Brillouin-Wigner perturbation
series (\ref{1d-1phonon}) which reproduces the general behavior of the dispersion curve.
As we already know, its main disadvantage is the lack of accuracy, especially in calculating
the value of the continuum edge. Thus, we propose to remedy this weak point
introducing correction terms "by hand".

First of all, we replace the first
order expression for the ground-state energy (which equals
$-\alpha \pi/2$ in this case) by its exact value $E(\alpha,0)$.
Then we also replace the
 approximate edge point which is unity by its exact value
$1+E(\alpha,0)$ in the propagator of the r.h.s. of
Eq.~(\ref{1d-1phonon}). Note that actually this step is not an approximation:
the energy can be arbitrarily shifted in the denominators of the Brillouin-Wigner
expansion, as was mentioned in Ref. \cite{lindemann}. This allows us
to keep the correct gap between the zone's bottom and the edge point.
Finally, we scale the total momentum $Q$
by the factor $b_1$ to obtain the correct effective mass behavior
of the type $E(\alpha,Q) \approx E(\alpha,0) + Q^2/m(\alpha)$ at
small $Q$. This way we arrive at our fitting formula:
\begin{eqnarray}
f&=&(b_1 Q)^2 + {E(\alpha,0)\over \sqrt{1+E(\alpha,0)-f}},
\nonumber \\
\label{interp1d} \\
b_1^2&=&{1\over
m(\alpha)}\left(1-{E(\alpha,0)\over 2}\right)\nonumber .
\end{eqnarray}
Later we apply a similar procedure to the cases $D=2, 3$.

At small $Q$ the correct parabolic behaviour is guaranteed by the very
construction of Eq.~(\ref{interp1d}). When $Q$ is large it leads
to the asymptotic behaviour
\begin{eqnarray}
f \approx 1+ E(\alpha,0) - {(E(\alpha,0))^2\over (b_1Q)^4} +
\dots  \label{asymp1d}
\end{eqnarray}
Thus, the dispersion curve approaches the correct continuum edge rather
fast.

The exact expressions for the polaron ground-state energy
$E(\alpha,0)$ and its effective mass $m(\alpha)$ are unknown
but they can be calculated numerically with the help of the
perturbation series. The first two terms of the latter were found
in Ref. \cite{peet2} and a few next terms were calculated by
Khomyakov\cite{khomyak}:
\begin{eqnarray}
E(\alpha,0) &=& -{\alpha\pi\over 2} -
0.060660\left({\alpha\pi\over 2}\right)^2
-0.00844437\left({\alpha\pi\over 2}\right)^3 \nonumber \\
&&-0.00151488\left({\alpha\pi\over 2}\right)^4,
\nonumber \\
m(\alpha)&=& 1+0.5 \left({\alpha\pi\over 2}\right) +
0.191942\left({\alpha\pi\over 2}\right)^2 \nonumber \\
&&+ 0.0691096 \left({\alpha\pi\over 2}\right)^3.
\label{eq1denergy}\end{eqnarray}

\begin{figure}[h]
%   \hspace*{-1.0cm}
%   \epsfysize=2.7in
   \includegraphics[width=3.7in]{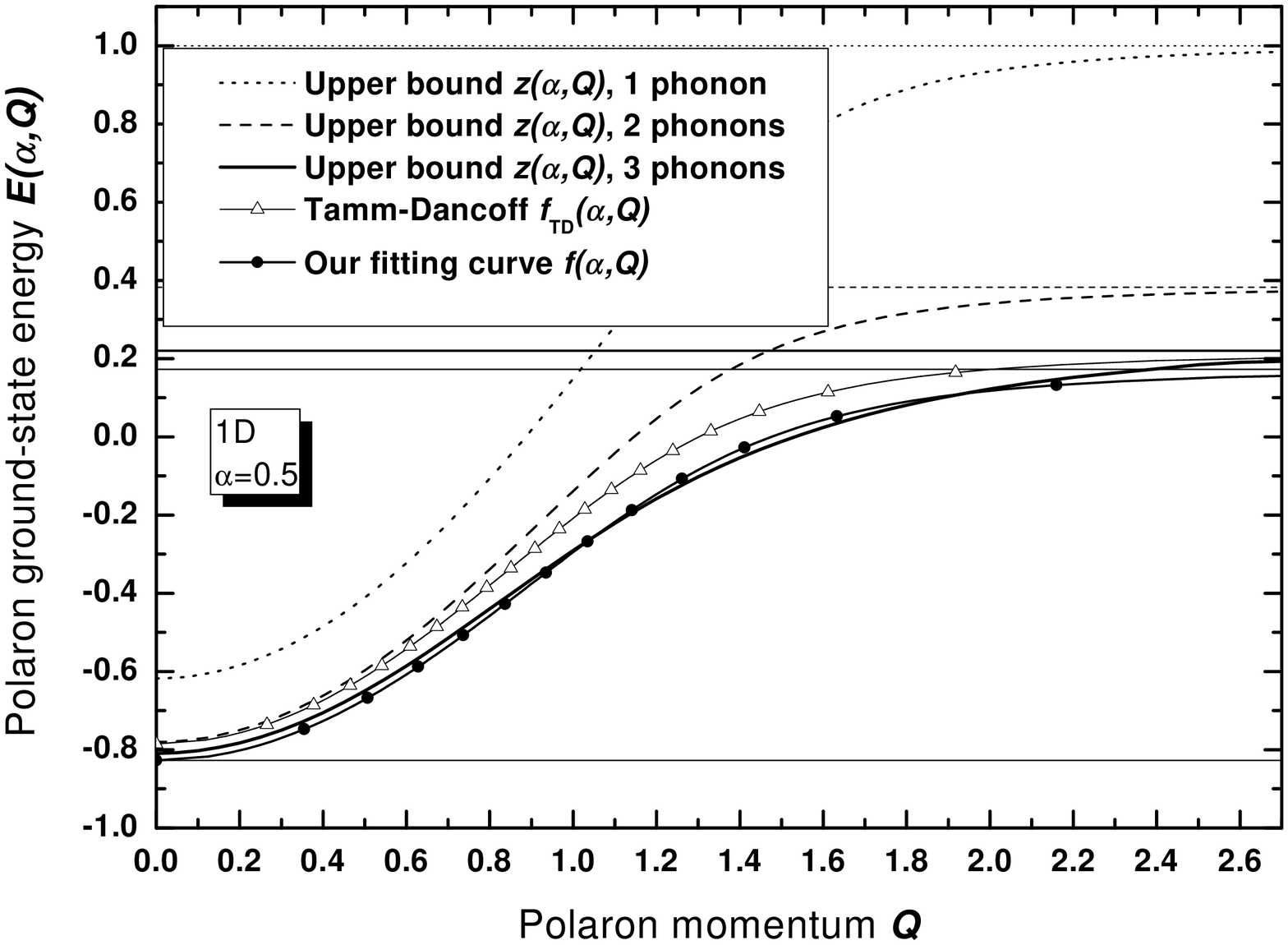}
%   \vspace*{-7.5cm}
\caption{Dispersion of 1D polaron for $\alpha=0.5$} \label{fig1a}
\end{figure}

The results for $\alpha=0.5$ are presented in Fig.~\ref{fig1a},
where numerically exact ground-state energy $E(\alpha,0)$ and the
edge-point $1+E(\alpha,0)$ are shown by solid thin straight lines.

Our fitting formula is close to the so-called {\it improved
Brillouin-Wigner perturbation theory} of Ref. \cite{lindemann}
which appears to be equivalent to the one-phonon Tamm-Dancoff
approximation \cite{whit}. Previously, this approximation was used
for polarons in $D=2, 3$ (the discussion and references will be
given later). With these ideas being applied to the polaron in 1D, we obtain the following expression:
\begin{eqnarray}
f_{TD}&=&Q^2 - {\alpha \pi /2 \over \sqrt{1-\alpha
\pi/2-f_{TD}}}. \label{tamm1d}
\end{eqnarray}
In comparison with our Eq.~(\ref{interp1d}) only the first
order in $\alpha$ is taken into account for the polaron ground
state energy at rest and the effective mass (which gives
$b_1=1$). The curve $f_{TD}$ is also plotted in Fig.~\ref{fig1a}.

Comparing with the three-phonon calculation one can notice that our
fitting curve provides us with an excellent result which is also better than the Tamm-Dancoff
approximation. This is not surprising because we used in our construction
the RSPT expansions to rather high orders which work quite well for the intermediate values of $\alpha$.
Thus, analogous fitting formulas will help us in the next sections
where it is not possible technically to perform
calculations within the three-phonon approximation to proceed to the region
of intermediate couplings.

\section{The 2D case}

With the one-phonon exchange taken into account (the same
Eqs.~(\ref{g}) and (\ref{1phonon}) taken for $D=2$) we arrive at
the equation for the upper bound:
\begin{equation}
z=Q^2 - {\alpha\over \sqrt{Q^2+1-z}}\ K\left( {Q^2\over Q^2+1-z}
\right). \label{2d-1phonon}\end{equation} If we put $z=Q^2$ in the
r.h.s. of Eq.~(\ref{2d-1phonon}), we get the expression for the
first order of the RSPT given by Eq.~(\ref{RSPT-2}).

\begin{figure}[h]
%   \hspace*{-1.0cm}
%   \epsfysize=2.7in
   \includegraphics[width=3.7in]{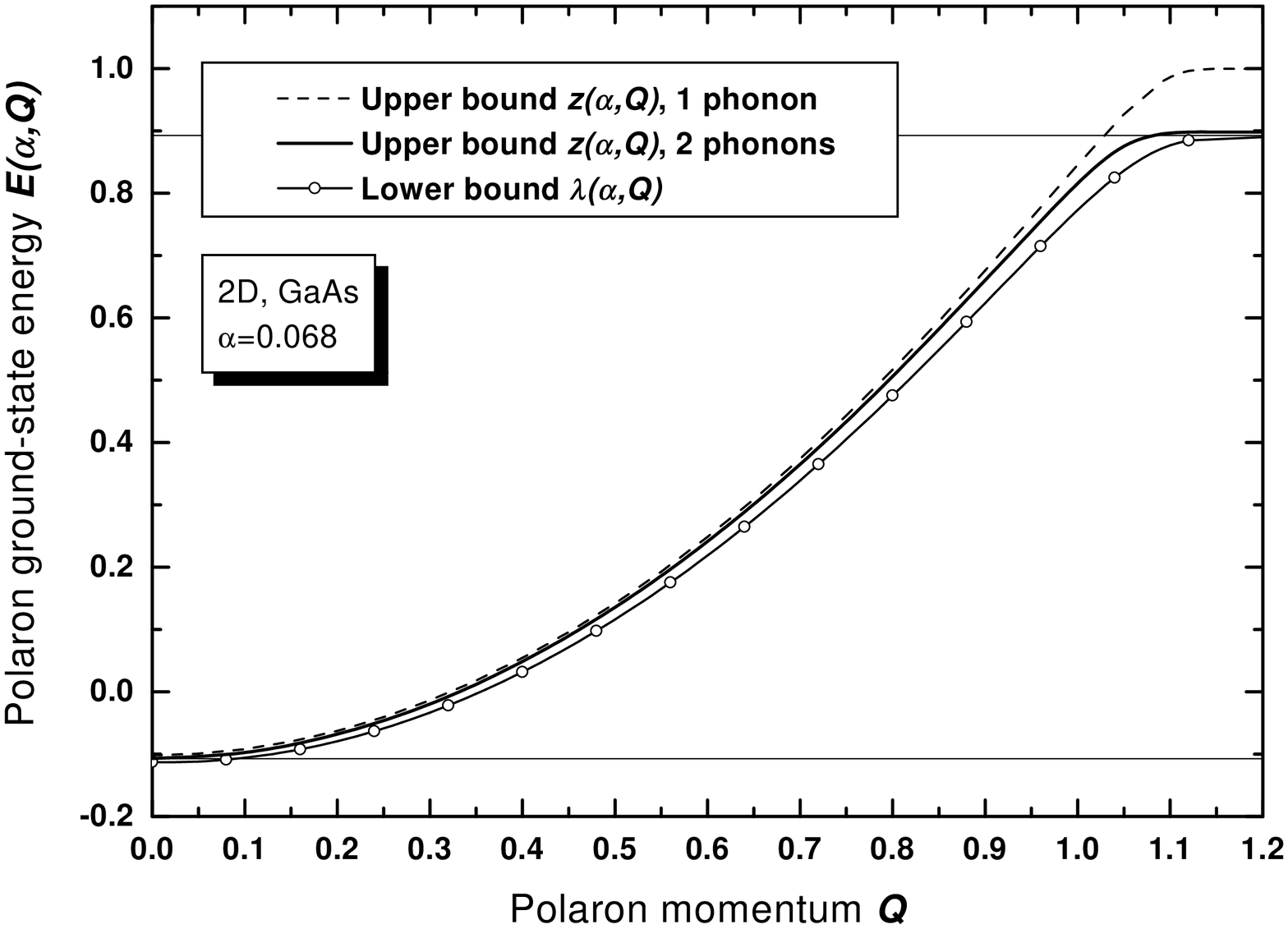}
%   \vspace*{-2.0cm}
\caption{Dispersion of 2D polaron in GaAs} \label{fig2}
\end{figure}

The results obtained with this formula and with the two-phonon
exchange as well as the lower bound are presented in
Fig.~\ref{fig2} for $\alpha=0.068$. Numerical
calculations for the three-phonon exchange are very complicated
and, therefore, will not be performed in this case. However, for
such a small value of $\alpha$ the gap between the upper and
lower bounds is very narrow.

To tackle intermediate values of $\alpha$, we apply the same idea to construct
a fitting formula which reads in 2D as follows:
\begin{eqnarray}
f&=&(b_2 Q)^2 +{2\over \pi} {E(\alpha,0)\over \sqrt{(b_2
Q)^2+1+E(\alpha,0)-f}} \nonumber \\
&&\times K\left( {(b_2 Q)^2\over (b_2
Q)^2+1+E(\alpha,0)-f} \right),\nonumber \\
\label{eq2dfit} \\
 b_2^2&=&{1\over m(\alpha)}\left(1-{E(\alpha,0)\over
 4-E(\alpha,0)}\right).\nonumber
\end{eqnarray}

Here again the correct parabolic behaviour at small $Q$ is
reproduced by the very construction of the fitting formula.
The asymptotic behavior of the complete elliptic integral $K(m)
\sim - (1/2) \ln (1-m / 16)$ when $m \to 1$ leads to the
following asymptotics of the dispersion curve at large $Q$:
\begin{eqnarray}
f \approx 1+ E(\alpha, 0)  - 16(b_2Q)^2 \exp [- {\pi (b_2Q)^3
\over |E(\alpha,0)| } ] + \dots
 \label{asymp2d}\end{eqnarray}
Thus, the dispersion curve approaches the continuum edge extremely fast.

The expressions for the $2D$-polaron ground-state energy and
effective mass can be taken from Ref. \cite{devr,selugin}:
\begin{eqnarray}
E(\alpha,0)&=& -{\pi\over2}\alpha-0.063974 \alpha^2, \nonumber \\
m(\alpha)&=& 1+{\pi\over 8} \alpha+0.127235 \alpha^2.
\label{eq2denergy}
\end{eqnarray}

\begin{figure}[h]
%   \hspace*{-1.0cm}
%   \epsfysize=2.8in
   \includegraphics[width=3.7in]{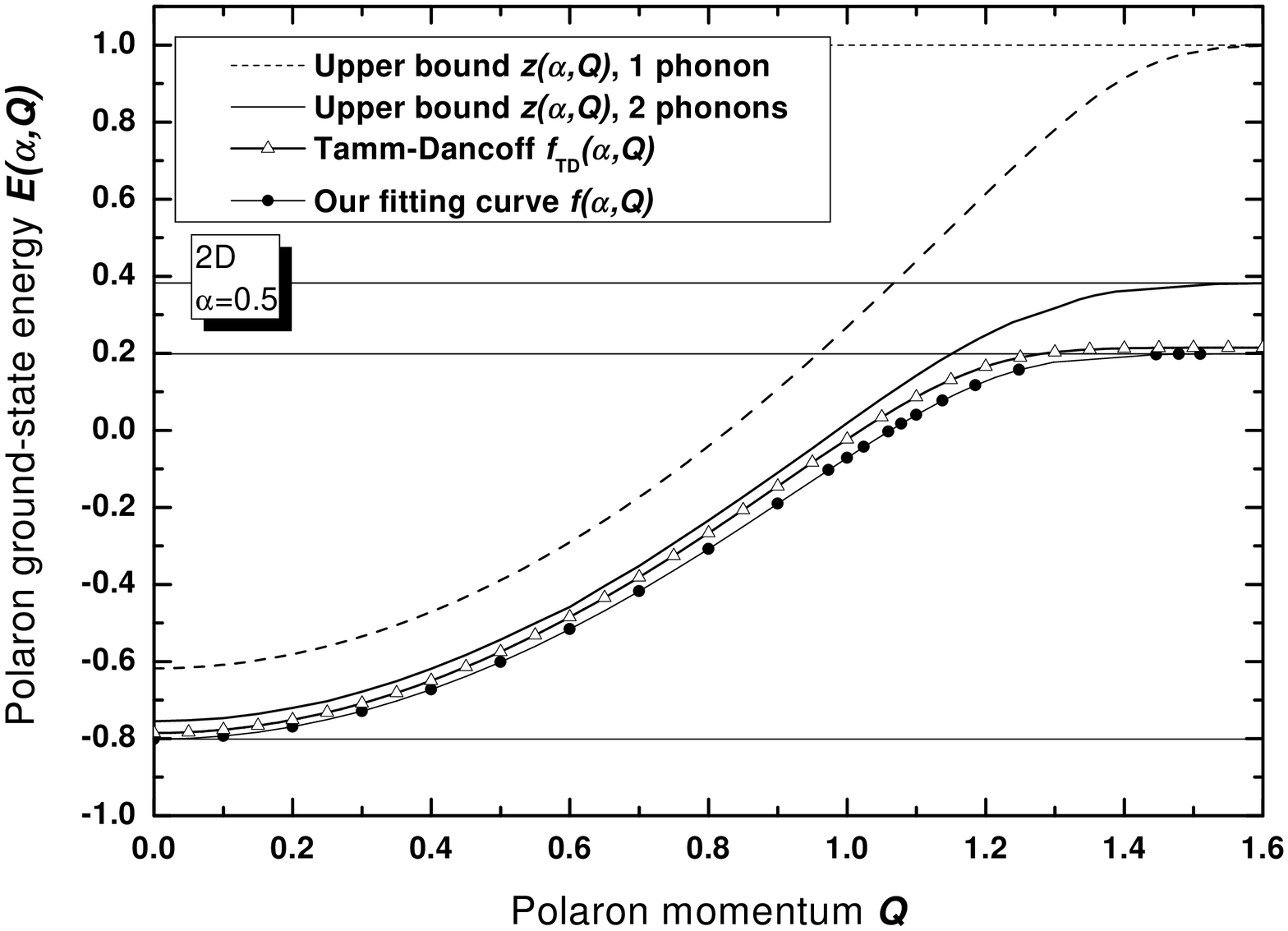}
%   \vspace*{-7.5cm}
\caption{Dispersion of 2D polaron for $\alpha=0.5$} \label{fig2a}
\end{figure}

The Tamm-Dancoff approximation for the 2D polaron was calculated
by the Antwerp group\cite{warmenbol}. Again, the formula is very
similar to our Eq.~(\ref{eq2dfit}), where $b_2$ is replaced by
unity and $E(\alpha,0)$ is put equal to $-\alpha\pi/2$:
\begin{eqnarray}
f_{TD}&=&Q^2 - {\alpha \over \sqrt{
Q^2+1-\alpha \pi/2-f_{TD}}} \nonumber \\
&&\times K\left( {Q^2\over Q^2+1-\alpha \pi/2-f_{TD}} \right).
\label{tamm2d}
\end{eqnarray}

The numerical results are shown in Fig.~\ref{fig2a}.

\section{The 3D case}

With Eqs.~(\ref{g}) and (\ref{1phonon}) the result of the
one-phonon approximation for the bulk polaron ($D=3$) reads as
follows:
\begin{equation}
z=Q^2 -{\alpha \over Q}\arcsin {Q \over \sqrt{Q^2+1-z}}.
\label{3d-1phonon}\end{equation} This equation appeared at first
in Ref.\cite{zienau}. Putting $z=Q^2$ in the r.h.s. of
Eq.~(\ref{3d-1phonon}) we get the expression for the first order
of the RSPT given by Eq.~(\ref{RSPT-3}).

Being expanded in powers of the coupling constant this
reproduces at $Q=0$ the exact results to the first order in $\alpha$:
\begin{equation}
z(\alpha,0)= -\alpha + {\cal O}(\alpha^2).
\end{equation}
For the edge point we have again $z_c=1$. Inserting this value
into Eq.~(\ref{3d-1phonon}) we obtain the equation for the value
$Q_c$ at which the edge point is reached:
\begin{equation}
Q_c(Q_c^2-1)=\alpha{\pi\over 2}. \label{1phonon4qc}\end{equation}

In this approximation the curve $z(\alpha,Q)$ approaches the edge
point $z(\alpha,Q_c)=1$ as an inverse parabola:
\begin{equation}
z(\alpha,Q)\approx 1-{Q_c^2(3Q_c^2-1)\over \alpha^2}(Q_c-Q)^2
 \label{eq_edge}\end{equation}
 at $Q \approx Q_c, Q < Q_c$.

\begin{figure}[h]
%  \hspace*{-1.0cm}
%  \epsfysize=2.7in
   \includegraphics[width=3.7in]{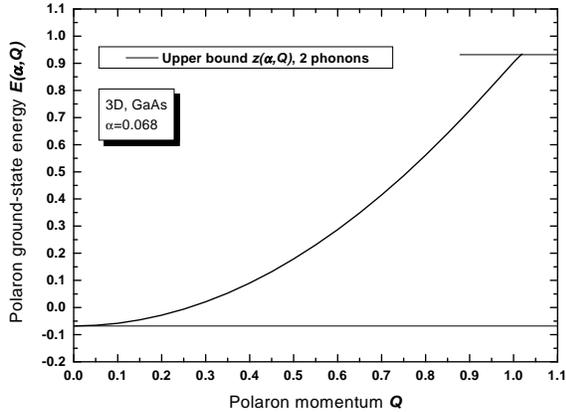}
  % \vspace*{-7.5cm}
\caption{Dispersion of 3D polaron in GaAs} \label{fig3}
\end{figure}

\begin{figure}[h]
 % \hspace*{-1.0cm}
%  \epsfysize=2.7in
   \includegraphics[width=3.7in]{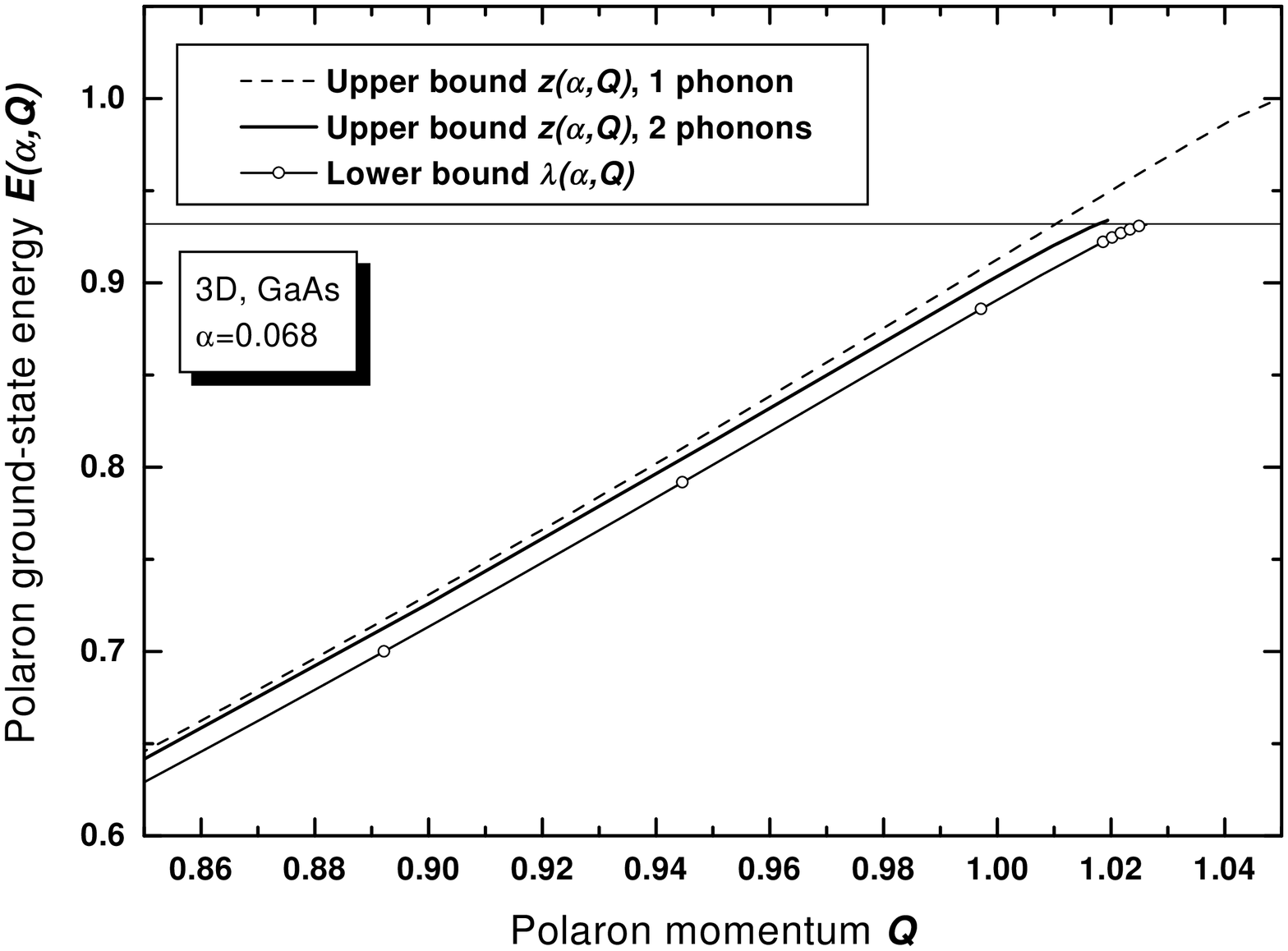}
 \vspace*{-0.5cm}
\caption{Dispersion of 3D polaron in GaAs near the edge point}
\label{fig4}
\end{figure}

The curve for the upper bound which we calculated within the
two-phonon approximation for GaAs is shown in Fig.~\ref{fig3}. In
the whole range of momentum $Q$ it is almost undistinguishable
from the lower and upper bounds within the one-phonon
approximation. Thus, it can be considered as a (numerically) exact
result. All three curves are presented near the edge point in
Fig.~\ref{fig4}. The curvature of the dispersion near the
edge point is very large at small $\alpha$, as it follows from Eq.~(\ref{eq_edge}),
and the inverse parabola described by this equation can hardly be seen in these plots.

Again the gap between the edge point and the zone's bottom
is larger than the phonon energy (unity) in the one-phonon approximation.
There is no physical reasons for that, and this disadvantage was removed in Ref.~\cite{whit}.
However, the authors obtained wrong
weak-coupling expansions for the polaron energy and effective mass at small momenta.
A modification of Eq.~(\ref{3d-1phonon}) was given by
Klochikhin\cite{klochikhin} who studied the 3D polaron dispersion
in the scope of the perturbation theory and took into account
two-phonon amplitudes. This way he arrived at the following equation:
\begin{equation}\label{kloch}
f_{TD}=Q^2 -{\alpha \over Q}\arcsin {Q \over \sqrt{Q^2+1-\alpha
-f_{TD}}},
\end{equation}
which coincides with the Tamm-Dancoff approximation \cite{warm2}.
This way he obtained the value $E_c \approx 1-\alpha$ for
the continuum edge in the first order in $\alpha$ which is reached at $Q_c=1.16$. The correct
energy shift was the advantage of these calculations in comparison with previously made in Ref.~\cite{whit}.

For intermediate values of $\alpha$ (we present here examples with
$\alpha=0.5, 1$) the two-phonon exchange contribution does not
provide us with impressive results. The three-phonon contribution
requires a very large amount of computational work and, therefore,
will not be given here. Besides, such values of $\alpha$ are
outside the domain admissible for the lower bound, although the
polaron is still in the weak-coupling regime. So we again construct
our fitting formula which now takes the form:
\begin{eqnarray}
f&=&(b_3 Q)^2 +{E(\alpha,0) \over b_3 Q}\arcsin {b_3 Q \over
\sqrt{(b_3 Q)^2+1+E(\alpha,0)-f}}, \nonumber \\
 b_3^2&=& {1\over m(\alpha)} \left(1-{E(\alpha,0)\over
6-2E(\alpha,0)}\right). \label{eq3dfit} \end{eqnarray} Note that
at $b_3=1$ and $E(\alpha,0) = -\alpha$ this equation reduces to
Eq.~(\ref{kloch}).

The bulk polaron ground state energy and its effective mass are
given by the known perturbation series (the second order in
$\alpha$ was found in Ref. \cite{roseler} and the third order was
calculated in Ref. \cite{smond,smond2}:
\begin{eqnarray}
E(\alpha, 0)  &=&  -\alpha - 1.59196 (\alpha /10)^2 - 0.80607
(\alpha /10)^3,  \nonumber \\
m(\alpha) &=&  1+\alpha/6+ 2.36276
(\alpha / 10)^2. \label{eq-perturb}
\end{eqnarray}

The maximal value $Q_c$ of the polaron momentum can be found from
the fitting formula (\ref{eq3dfit}) if we put
$f=1+E(\alpha,0)$. Then we arrive at the equation for $Q_c$:
%% which replaces Eq.~(\ref{1phonon4qc}) :
\begin{eqnarray}
&& Q_c = q_c/b_3, \nonumber \\
&& q_c \left(q_c^2-1-E(\alpha,0)\right)=-{\pi \over 2}E(\alpha,0).
 \label{eq4qc}\end{eqnarray}

The fitting dispersion curve $f(\alpha,Q)$ approaches the edge
point also as an inverse parabola  at $Q \approx Q_c, Q < Q_c$:
\begin{eqnarray}
f(\alpha,Q) &\approx & 1+E(\alpha,0)  \nonumber \\
& - &{q_c^2(3q_c^2-1-E(\alpha,0))^2\over
E(\alpha,0)^2}(q_c-q)^2,
 \label{eq_edge2}\end{eqnarray}
where $q= b_3Q$.
%% This equation is the improved version of Eq.~(\ref{eq_edge}).

\begin{figure}[h]
 %  \hspace*{-1.0cm}
%   \epsfysize=2.7in
   \includegraphics[width=3.7in]{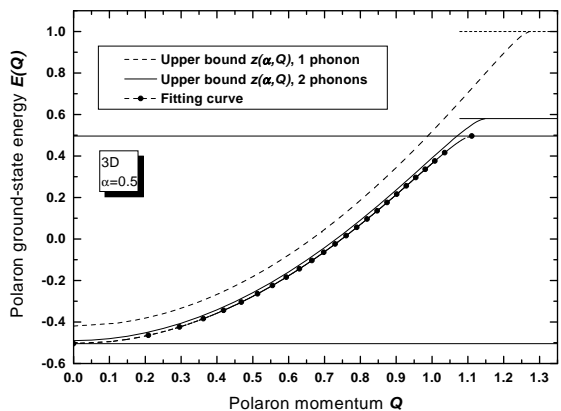}
%   \vspace*{-7.5cm}
\caption{Dispersion of 3D polaron for $\alpha = 0.5$} \label{fig5}
\end{figure}

\begin{figure}[th]
%   \hspace*{-1.0cm}
%   \epsfysize=2.7in
   \includegraphics[width=3.7in]{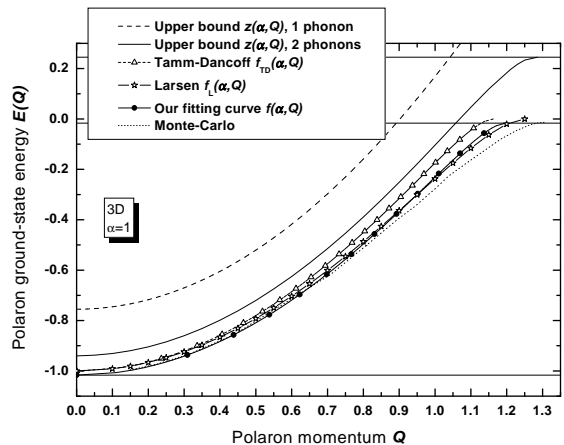}
%   \vspace*{-3.0cm}
\caption{Dispersion of 3D polaron for $\alpha = 1$} \label{fig6}
\end{figure}

This way we found $Q_c = 1.02$ for $\alpha=0.068$, $Q_c = 1.11$
for $\alpha=0.5$, and $Q_c=1.20$ for $\alpha=1$. The fitting
curves obtained are shown in Figs.~\ref{fig5} and \ref{fig6}
together with one- and two-phonon calculations. Being applied at
$\alpha=0.068$ this procedure leads practically to the same
results as direct calculations.

The results of the Monte-Carlo
calculations\cite{prokofiev,prokofiev2} are shown also in
Fig.~\ref{fig6}. The relation $k=Q \sqrt{2}$ is used between our
momentum $Q$ and their momentum $k$ (free electron energy reads
as follows: $E=Q^2=k^2/2$). We see excellent agreement between
our results almost in  the whole range of momentum except for the
vicinity of the edge point. The Monte-Carlo calculations give the
value $Q_c^{MC}
 \approx 1.285$ while our fitting formula gives $Q_c^{fit} =
1.20$. This discrepancy of 8\% is responsible for the deviation of
these curves. The authors of Refs. \cite{prokofiev,prokofiev2}
reported that their calculations were "numerically exact" and the
error bars were smaller than the size of points at the plot.
Previously, \cite{kalina} we criticized these statements,
in particular, because  the results obtained
give the wrong coefficient even in the second order in $\alpha$
for the polaron ground state energy weak-coupling expansion.
As we could see above, the correct second order calculations
are crucial for the adequate description of the dispersion near the edge point.

The polaron dispersion $f(\alpha, Q)$ was also calculated by
Larsen \cite{larsen} who combined for his variational ansatz the
one-phonon Tamm-Dancoff approximation and the Lee-Low-Pines
transformation \cite{lee}:

\begin{eqnarray}
{f_L +\alpha -Q^2 \over f_L +\alpha + Q^2} = {2\alpha \over \pi}
F(\alpha,Q), \nonumber \\
F(\alpha, Q) = \int_0^{\infty} dk {k^2\over (1+k^2)^2}\cdot  \nonumber \\
\cdot \int_{-1}^{1} d\xi {\xi^2 \over f_L +\alpha -
1-Q^2-k^2+2Qk\xi}.
 \label{larsen}\end{eqnarray}

Both the Tamm-Dancoff approximation (\ref{kloch}) and the
calculations with Eq.~(\ref{larsen}) are shown in Fig.~\ref{fig6}.
Our curve is very close to Larsen's one although they lead to
different values of the edge point momentum ($Q_c^{Lar} \approx
1.25$). On the other hand, Larsen's curve tends to a higher (and wrong) value
of the polaron energy at the edge point and this is the main source of the
discrepancy in calculating $Q_c$. However, if we recall that the correct edge point value
$1+E(\alpha, 0)$ is the rigorous upper bound for the dispersion (see Eq.~(\ref{rigorous})),
we have to find the momentum at which Larsen's curve reaches it. This way we found
practically the same $Q_c \approx 1.20$ as given by our fitting formula.
Thus, we arrive at the same results for $Q_c$ although Larsen's Eq.~(\ref{larsen}) and our
Eq.~(\ref{eq3dfit}) do not look similar and were obtained in
different ways. This gives hope that the value of $Q_c =1.20$ found for $\alpha=1$
fits the exact one quite well.

%%One can also estimate roughly the limiting value $Q_c$ using the calculations in one- and
%%two-phonon approximations. At $\alpha=1$ the end points of the dispersion have
%%coordinates $Q^{1\,ph}_c=1.4447, E^{1\,ph}=1$ and $Q^{2\,ph}_c=1.2854,
%%E^{2\,ph}=0.2451$. At each $n$ the values $Q^{n\,ph}$ becomes closer to each other and to
%%the exact value $Q_c$. All these end points lay at some curve. If we approximate the
%%latter by a straight line, connecting the end points calculated
%%within one- and two-phonon approximations,
%%we can find the momentum $Q_c^{sl}$ at which this line intercepts the correct asymptote
%%$E_c=1+E(\alpha,0)=-0.016726$ . This happens at $Q_c^{sl}=1.23$. Of course, this is
%%a rather crude estimate, but surprisingly this value is rather close (2.5\%) to that by
%%Larsen and by the present authors. Similar results obtained in so
%%different approximations give hope that they fit the exact one
%%quite well.

\section{Conclusions}

We have presented here the upper and lower bounds for the polaron
dispersion in $D=1,2,3$. At small values of the
electron-phonon coupling constant $\alpha$ the gap between our
estimates is so narrow that we report in reality numerically exact results.

For intermediate values of $\alpha \approx 1$ we proposed the fitting
formula which is simple to use in numerical calculations. Its
validity is proved by comparing with the direct three-phonon
calculations for the $1D$-polaron. For $D=2, 3$ the comparison with
the previously obtained results is also made, which allows us
to conclude that the proposed formula fits the exact dispersion quite
well in all space dimensions.

%%Natural question arises if we can apply our fitting
%%formulae (\ref{interp1d}), (\ref{eq2dfit}) and (\ref{eq3dfit}) to
%%the strong coupling limit of very large $\alpha$. At first glance
%%nothing prevents us from replacing the weak-coupling expressions
%%for the ground-state energy $E(\alpha,0)$ and the effective mass
%%$m(\alpha)$ by their strong-coupling analogs to arrive at
%%appropriate interpolation formulae. But let us remind that the
%%latter have been constructed on the basis of the first order
%%Brillouin-Wigner perturbation theory. Thus, we cannot be sure
%%that the general behavior of the dispersion is reproduced
%%adequately at large $\alpha$.
%%
%%On the other hand, Larsen's approach which gives rather similar
%%results was suggested to cover intermediate values of the
%%coupling constants. Note also that despite countless revocations,
%%no self-trapping transition does exist at large values of
%%$\alpha$, as was proven by Gerlach and L\"owen \cite{gerl2}. This
%%means that the dispersion curve should transform smoothly with
%%increasing $\alpha$. Although the study of the strong-coupling
%%limit is beyond the scope of the present paper, we would not be
%%surprised if the above-mentioned minimal changes in our
%%interpolation formulae will lead to satisfactory results.

\section*{Acknowledgments}

We thank Dr. Frank Kalina in collaboration with whom we
calculated lower bounds for polaron in different spatial
dimensions. We are grateful to Prof. Fran\c{c}ois Peeters and
Prof. Vladimir Fomin for their
attention to the paper and valuable remarks. M.A.S. thanks University of Dortmund for the
hospitality during his visits to Germany. The study was performed
with the financial support of Deutsche Forschungsgemeinschaft and
the Heisenberg-Landau program.


\begin{thebibliography}{99}
\bibitem{zienau} H. Fr\"ohlich, H. Pelzer, and S. Zienau, Phil.
Mag. {\bf 41}, 221 (1950).
\bibitem{whit}  G. Whitfield, and R. D. Puff, Phys. Rev. {\bf{139}}, A338 (1965).
\bibitem{appel} J. Appel, Polarons, in: Solid State Physics (Advances in Research and
Applications), {\bf 21}, ed. by F. Seitz, D. Turnbull and H.
Ehrehreich (Academic Press, New York, 1968), p. 193.
\bibitem{lowen} B. Gerlach, and H. L\"{o}wen, Rev. Mod. Phys. {\bf
63}, 63 (1991).
\bibitem{spohn} H. Spohn, Ann. Phys. {\bf 175}, 278 (1987).
\bibitem{lee} T. D. Lee, F. E. Low, and D. Pines, Phys. Rev. {\bf
90}, 297 (1953); {\bf 91}, 227 (1953).
\bibitem{larsen} D. M. Larsen, Phys. Rev. {\bf 144}, 697 (1966).
D. M. Larsen, in: {\it Polarons in Ionic Crystals and Polar
Semiconductors}, ed. J. T. Devreese, (North-Holland, Amsterdam,
1972), p. 237.
\bibitem{warm2} P. Warmenbol, F. M. Peeters, and J. T. Devreese,
Phys. Rev. B {\bf 33}, 5590 (1986).
\bibitem{warmenbol} F. M. Peeters, P. Warmenbol, and J. T.
Devreese, Europhys. Lett. {\bf 3}, 1219 (1987).
\bibitem{lieb} E. Lieb, and K. Yamazaki, Phys. Rev. {\bf 111}, A
728 (1958). See also E. Lieb, and L. Thomas, Comm. Math. Phys. {\bf
183}, 511 (1997); Erratum: Comm. Math. Phys. {\bf 188}, 499
(1997).
\bibitem{gerl3} B. Gerlach, and F. Kalina Phys. Rev. B {\bf{60}}, 10886 (1999).
\bibitem{peet1} F. M. Peeters, Wu Xiaoguang, and J. T. Devreese, Phys.
Rev. B {\bf{33}}, 3926 (1986).
\bibitem{peet2} F. M. Peeters, and M. A. Smondyrev, Phys. Rev. B {\bf{43}},
                4920 (1991).
\bibitem{gerl2} B. Gerlach, and H. L\"owen, Phys. Rev. B {\bf{35}}, 4291 (1987).
\bibitem{kalina} B. Gerlach, F. Kalina, and M. A. Smondyrev, phys.
stat. sol. (b) {\bf 237}, 204 (2003).
\bibitem{spo2}  H. Spohn, J. Phys. A {\bf{21}}, 1199 (1988).
\bibitem{karth} E. Kartheuser, in: {\it Polarons in Ionic Crystals and Polar
Semiconductors}, ed. J. T. Devreese, (North-Holland, Amsterdam,
1972), p. 717.
\bibitem{feynman} R. P. Feyman. Phys. Rev. {\bf 84}, 108 (1951).
\bibitem{smond2} O. V. Selyugin, and M. A. Smondyrev, phys. stat. sol. (b) {\bf
155}, 155 (1989).
\bibitem{larsen2} D. Larsen, Phys. Rev. {\bf 187}, 1147 (1969).
\bibitem{klochikhin} A. A. Klochikhin, Fiz. Tverd. Tela {\bf 21},
3077 (1979) [Sov. Phys. - Solid State {\bf 21}, 1770 (1980)].
\bibitem{smond} M. A. Smondyrev, Teor. Mat. Fiz. {\bf 68}, 29 (1986) [Theor. Math. Phys. {\bf 68}, 653
(1987)].
\bibitem{lindemann}G. Lindemann, R. Lassnig, W. Seidenbusch, and
E. Gornik, Phys. Rev. B {\bf 28}, 4693 (1983).
\bibitem{khomyak}P. A. Khomyakov, Phys. Rev. B {\bf 63}, 153405-1
(2001).
\bibitem{devr} Wu Xiaoguang, F. M. Peeters, and J. T. Devreese, Phys. Rev.
B {\bf{31}}, 3420 (1985).
\bibitem{selugin} O. V. Seljugin, and M. A. Smondyrev, Physica
{\bf{142A}}, 555 (1987).
\bibitem{roseler} J. R\"oseler, phys. stat. sol. {\bf 25}, 311
(1968).
\bibitem{prokofiev} N. V. Prokof'ev, and B. V. Svistunov, Phys. Rev. Lett. {\bf
81}, 2514 (1998).
\bibitem{prokofiev2} A. S. Mischenko,  N. V. Prokof'ev, A. Sakamoto, and B. V. Svistunov, Phys. Rev. B {\bf
62}, 6317 (2000).


\end{thebibliography}
\end{document}